\documentclass[fleqn,a4paper,12pt]{article}
\usepackage{amssymb}
\usepackage{amsmath}
\usepackage[dvips]{graphicx}
\usepackage{placeins}
\usepackage{lscape}
\usepackage{a4}
\usepackage{epsfig}

\newcommand{\ri}{{\mathrm i}}
\newcommand{\p}{\partial}
\newcommand{\bea}{\begin{array}}
\newcommand{\eea}{\end{array}}

\catcode`@=11 \long
\def\@caption#1[#2]#3{\par\addcontentsline{\csname
ext@#1\endcsname}{#1} {\protect\numberline{\csname
the#1\endcsname}{\ignorespaces #2}} \begingroup \small
\@parboxrestore \@makecaption{\csname fnum@#1\endcsname}
{\ignorespaces #3}\par \endgroup} \catcode`@=12

\newcommand{\la}{\label}

\catcode`@=11 \long
\def\@caption#1[#2]#3{\par\addcontentsline{\csname
ext@#1\endcsname}{#1} {\protect\numberline{\csname
the#1\endcsname}{\ignorespaces #2}} \begingroup \small
\@parboxrestore \@makecaption{\csname fnum@#1\endcsname}
{\ignorespaces #3}\par \endgroup} \catcode`@=12

\begin{document}

\allowdisplaybreaks
 \begin{titlepage} \vskip 2cm

\begin{center} {\Large\bf  New exactly solvable systems with Fock symmetry} \footnote{E-mail:
{\tt nikitin@imath.kiev.ua} } \vskip 3cm {\bf {A. G. Nikitin }
\vskip 5pt {\sl Institute of Mathematics, National Academy of
Sciences of Ukraine,\\ 3 Tereshchenkivs'ka Street, Kyiv-4, Ukraine,
01601\\}}\end{center}
\vskip .5cm \rm
\begin{abstract}
New superintegrable systems are presented which, like the
Hydrogen atom, possess a dynamical symmetry w.r.t. algebra o(4). One of them simulates a neutral fermion with non-trivial dipole moment, interacting with the external e.m. field. This system is presented in both non-relativistic and relativistic formulations. Another recently discovered system (see arXiv:1208.2886v1) is non-relativistic and includes the minimal and spin-orbit interaction with the external electric field. It is shown that all the considered systems are shape invariant. Applying this quality, these systems are integrated  using the tools of SUSY quantum mechanics.
\end{abstract}
\end{titlepage}
\section{Introduction\label{intro}} The Hydrogen atom (HA) is one of the
most important  systems of   quantum mechanics. This is a perfect physical model with
a  nice symmetry. Namely, in addition to its transparent invariance with
respect to the rotation group, the HA possesses a hidden (Fock
\cite{Fock}) symmetry w.r.t. group O(4) whose generators are the
orbital momentum and Runge-Lenz vector \cite{barg}. Moreover, this
system is also supersymmetric and can be solved algebraically using
tools of SUSY quantum mechanics \cite{gen}.

At the best of my knowledge,   until the first version of this paper appeared in ArXiv-1205.3094v1 there were only two known generalizations
of the Runge-Lenz vector for 3d QM systems with spin. The first of them was discovered by Johnson and Lippman as far back as in 1950 \cite{John} as a hidden symmetry of the relativistic Dirac equation with
Coulomb potential. A contemporary treatment of the Johnson-Lippman
constant of motion in arbitrary dimensional space is presented in
\cite{japona}. The other generalization was proposed in 1985 by D'Hoker and L.
Vinet who proved that the Schr\"odinger-Pauli equation for electron
interacting with the dyon field admits a vector integral of motion depending on spin \cite{Hock}.
However, analogy of these symmetries with the Runge-Lenz
vector is rather poor. Indeed,  the Johnson-Lippman
integral of motion is a scalar which is decoupled to three vector components only in the non-relativistic limit. The constant of motion found in \cite{Hock} includes the same higher derivative terms as the Runge-Lenz vector, but it does not generate the dynamical symmetry w.r.t. group O(4).

In very recent paper \cite{wint2} superintegrable QM systems with spin-orbit interaction and second-order integrals of motion are classified. One of these systems which includes the generalized Runge-Lenz vector is shortly discussed below in Section 6.

Notice that the spin dependent Runge-Lenz vector was discovered also
for  for spinning Taub-NUT space \cite{Vin}, but it was done on the
pseudoclassical level.

In the present paper a new QM system with spin 1/2,  which admits a
hidden  symmetry w.r.t. group O(4), is discussed. Its integrals of motion are the
total orbital momentum and the generalized Runge-Lenz vector
depending on spin. Like the HA, this system appears to be
supersymmetric also, which makes it possible  to find its exact solutions using
the regular SUSY approach.

Mathematically, the models presented in the following are  interesting new examples of superintegrable  and supersymmetric systems with spin, which are exactly solvable. These systems include shape invariant matrix potentials, which are particular cases of such potentials classified recently in \cite{NK1} ,  \cite{NK2}. In addition, these examples corroborate the conjecture that all maximally superintegrable
systems are exactly solvable \cite{wint3} (see  \cite{wint4}, \cite{ford}, \cite{kal}  for discussion) and present a new field for studying the relations between supersymmetry and superintegrability.

Physically, the models discussed in sections 3-5 simulate a neutral fermion  with a non-trivial dipole moment (e.g., the neutron), interacting
with the external field. Moreover, a relativistic version of this model is presented also.   lt is shown that the neutron can be trapped by the specific external fields.   Potentially, these systems and their exact solutions  can have a wide spectrum of applications, starting with using them as tutorial examples and ending with relevance  to the problems of security of nuclear reactors. In any case, the  superintegrable systems like the HA and isotropic harmonic oscillator are extremely important in physical applications. Thus we can hope that, to some degree, it would be the case for the systems discussed in the present paper.

\section{Symmetries of the Hydrogen atom}

Let us remind the main symmetry properties of the HA which will be used as
a standard for construction of the model with spin.

 The  Hamiltonian of the HA looks as follows:
\begin{gather}H=\frac{p^2}{2m}-\frac{q}{x}\la{H1}\end{gather}
where $p^2=p_1^2+p_2^2+p_3^2$, $p_1=-i\frac{\p}{\p x_1}$,
$x=\sqrt{x_1^2+x_2^2+x_3^2},\ q>0$.

Operator (\ref{H1}) is manifestly invariant w.r.t. rotation group
O(3) and so it commutes with the angular momentum vector:
\begin{gather}\la{L}{\bf L}={\bf x}\times {\bf p}. \end{gather}
In addition, it commutes with the Runge-Lenz vector
\begin{gather}\la{R}{\bf R}=\frac1{2m}({\bf p}\times {\bf L}-
{\bf L}\times{\bf p})+{\bf x}V \end{gather} where
$V=-\frac{q}{x}$.

Components of vectors (\ref{L}) and (\ref{R}) satisfy the following
commutation relations
\begin{gather}\la{cr}\begin{split}&[L_a,L_b]=\ri\varepsilon_{abc}L_c,\quad
[R_a,L_b]=\ri\varepsilon_{abc}R_c,\\&[R_a,R_b]=-\frac{2\ri}{m}\varepsilon_{abc}L_c
H.
\end{split}\end{gather}
On the set of eigenvectors of Hamiltonian $H$ algebra (\ref{cr}) is
isomorphic to the Lie algebra of group O(4).

Thus there are the six integrals of motion for Hamiltonian (\ref{H1}) (we do not take into account the Hamiltonian itself). However, we can select maximally  four such integrals which
are algebraically independent and include a pair of commuting representatives. This means that the
HA is a maximally superintegrable system: the number of its
algebraically independent constants of motion including the Hamiltonian is equal to $2n-1$, i.e., is maximal for a
system with $n=3$ degrees of freedom.

The other (discrete) symmetry of Hamiltonian (\ref{H1}) is the space
reflection $P$. Indeed, this Hamiltonian is transparently invariant
w.r.t. the change ${\bf x}\to-{\bf x}$.

In addition, operator (\ref{H1}) admits a hidden supersymmetry
\cite{gen}. In other words, its radial component is shape invariant
with respect to the special Darboux transformation which will be
specified in Section 4.

\section{Runge-Lenz vector for fermions}
Operators (\ref{L}) and (\ref{R}) by construction correspond to a
spinless system. A straightforward way to generalize them for the case of
a fermion system is to change orbital momentum $L$ by the total angular
momentum:
\begin{gather}\la{J}\begin{split}&{\bf L}\to{\bf J}={\bf L}+{\bf
S},\\& {\bf R}\to{\hat {\bf R}}=\frac1{2m}({\bf p}\times {\bf J}-
{\bf J}\times {\bf p})+{\bf x}\hat V\end{split}\end{gather}
where $\bf S=\frac12\mbox{\boldmath $\sigma$} $ is the spin vector
and $\mbox{\boldmath $\sigma$} $  is the matrix vector whose
components are Pauli matrices. Potential $\hat V$ should commute
with $\bf J$ and is in general spin dependent.

Operators (\ref{J}) should satisfy relations (\ref{cr}) where ${
L_a}\to  J_a, \ R_a\to{\hat R}_a$:
\begin{gather}\la{CR}\begin{split}&[J_a,J_b]=\ri\varepsilon_{abc}J_c,\quad
[\hat R_a,J_b]= \ri\varepsilon_{abc}\hat R_c,\\&[\hat R_a,\hat R_b]=-\frac{2\ri}{m}\varepsilon_{abc}J_c
H.
\end{split}\end{gather}
 and commute with a Hamiltonian
which we will search in the form
\begin{gather}\la{H2}H=\frac{p^2}{2m} +\hat V.\end{gather}

It can be verified by direct calculation that up to a constant
multiplier $\alpha$ we have the unique choice for $\hat V$:
\begin{gather}\la{V}\hat V=
\alpha\frac{\mbox{\boldmath $\sigma$} \cdot \bf
x}{x^2}.\end{gather}

Thus it is Hamiltonian (\ref{H2}) with "matrix Coulomb potential"
(\ref{V}) which admits the integrals of motion
(\ref{J}). This  Hamiltonian  is not invariant
w.r.t. space inversion ${\bf x}\to-{\bf x}$ since $\mbox{\boldmath
$\sigma$}$ is a pseudovector. However the space inversion is an
admissible transformation provided the wave function $\psi$
cotransforms in a non-standard way:
\begin{gather}\la{P}\psi(t,{\bf x})\to B\psi(t,-{\bf x})\end{gather}
where $B=\frac{{\mbox{\boldmath $\sigma$}}\cdot {\bf
L}+1}{|{\mbox{\boldmath $\sigma$}}\cdot {\bf L}+1|}$ is the
Biedenharn operator \cite{bied}. More exactly, the following
condition is satisfied:
\[BH({\bf p},{\bf x})B=H(-{\bf p},-{\bf x}).\]

 A natural question whether this Hamiltonian
succeeds another symmetry of (\ref{H1}), i.e., the shape invariance, is
discussed in the following section.

\section{Supersymmetry and exact solutions}

Consider the eigenvalue problem for Hamiltonian (\ref{H2}):
\begin{gather}\left(\frac{p^2}{2m}+
\alpha\frac{{\mbox{\boldmath $\sigma$}}\cdot{\bf
x}}{x^2}\right)\psi=E\psi.\la{EP}
\end{gather}
Introducing rescalled independent variables ${\bf r}=2m\alpha{\bf
x}$ it is possible to rewrite (\ref{EP}) in a more compact form:
\begin{gather}\left(-\Delta+\frac{{\mbox{\boldmath $\sigma$}}\cdot{\bf
r}}{r^2}\right)\psi=\varepsilon\psi\la{ep}\end{gather}
 where $\varepsilon
=\frac{E}{2m\alpha^2}$.

 Taking into account the invariance of equation
(\ref{EP}) w.r.t. the rotation group, it is convenient to rewrite it
in the spherical coordinates and expand $\psi$ via spherical spinors
$\Omega_{j,j-\lambda,\kappa}(\varphi,\theta)$:
\begin{gather}\psi=\frac1r\sum_{j,\lambda,\kappa} \psi_{j\lambda \kappa}(r)
\Omega_{j,j-\lambda,\kappa}(\varphi,\theta).\la{SS}\end{gather} Here
$j=\frac12,\ \frac32,\ \dots, \ \kappa=-j,\ -j+1,\ \dots\ j$, and
$\lambda=\pm\frac12$ are quantum numbers labeling eigenvalues of the
commuting operators ${\bf J}^2, \ {\bf L}^2$ and $J_3$:
\begin{gather*}\begin{split}&{\bf J}^2\Omega_{j,j-\lambda,\kappa}=
j(j+1)\Omega_{j,j-\lambda,\kappa},\\&{\bf
L}^2\Omega_{j,j-\lambda,\kappa}=(j-\lambda)(j-\lambda+1)\Omega_{j,j-\lambda,\kappa},
\\&J_3\Omega_{j,j-\lambda,\kappa}=k\Omega_{j,j-\lambda,\kappa}.\end{split}\end{gather*}
The explicit form of the spherical spinors is given by the following
formula \cite{biede}:
\begin{gather*}\Omega_{j,j-\frac12,\kappa}=\begin{pmatrix}\sqrt{\frac{j+
\kappa}{2j}}Y_{j-\frac12,\kappa-\frac12}\\\sqrt{\frac{j-
\kappa}{2j}}Y_{j-\frac12,\kappa+\frac12}\end{pmatrix},\quad
\Omega_{j,j+\frac12,\kappa}=\begin{pmatrix}-\sqrt{\frac{j-
\kappa+1}{2j+2}}Y_{j+\frac12,\kappa-\frac12}\\\sqrt{\frac{j+
\kappa+1}{2j+2}}Y_{j+\frac12,\kappa+\frac12}\end{pmatrix}\end{gather*}
where $ Y_{j\pm \frac12,
k\pm \frac12}$ are spherical functions.

 Substituting
(\ref{SS}) into (\ref{ep}) we come to the following equations
\begin{gather}{\cal H}_{j}\Phi_{j,\kappa}\equiv\left(-\frac{\p^2}{\p r^2}+
V_{j}\right)
\Phi_{j,\kappa}=\varepsilon\Phi_{j,\kappa}\la{rep}\end{gather} where
\begin{gather}\la{phi}\Phi_{j,\kappa}=\left(\begin{matrix}\psi_{j,-\frac12,\kappa}(r)\\
\psi_{j,\frac12,\kappa}(r)\end{matrix}\right)\end{gather} and
$V_{j}$ is the matrix potential:
\begin{gather}V_{j}=\left(j(j+1)+\frac14-\sigma_3\left(j+
\frac12\right)\right)\frac1{r^2}-\sigma_1\frac1r.
\la{pot}\end{gather}

Equations (\ref{rep}) appear to be supersymmetric since $V_{j}$
belongs to the list of shape invariant potentials classified in
\cite{NK1}, see Eq. (5.11) for $\kappa=\frac12,\ \mu=j$ there.
Indeed, Hamiltonian ${\cal H}_j$ can be factorized as
\begin{equation}\label{fac}{\cal H}_{j}=a_j^+a_j+c_j\end{equation} where
\begin{gather}\label{a+-}a_j=\frac{\partial}{\partial r}+W_j,\  \
a_j^+=- \frac{\partial}{\partial r}+W_j,\ \
c_j=-\frac{1}{4(j+1)^2}\end{gather} and $W_j$ is the {\it matrix
superpotential} \begin{gather}\label{SS1}
W_j=\left(\frac12\sigma_3-j-1\right)\frac1{r}
+\frac{1}{2(j+1)}\sigma_1.\end{gather} Moreover, Hamiltonians ${\cal
H}_j$ and ${\cal H}_{j+1}$ satisfy the following intertwining
relations
\begin{gather}\la{ir}{\cal H}_ja^+_j=a^+_j{\cal
H}_{j+1}.\end{gather}

Thus Eq. (\ref{rep}) can be easily solved using tools of SUSY
quantum mechanics. In fact it has been already done in \cite{NK1}.
The ground state $\Phi^0_{j,k}=\text{column}(\psi^0,\xi^0)$ should
solve the first order equation $a_j\Phi^0_{j,k}=0$. Thus, using
definitions (\ref{a+-}) we obtain:
\begin{gather}\label{gs}\Phi^0_{j,k}=c_k\begin{pmatrix} r^{j+\frac32}
K_{1}\left(\frac{r}{2(j+1)}\right)\\
r^{j+\frac32}K_0\left(\frac{r}{2(j+1)}\right)
\end{pmatrix}\end{gather}
where $K_1$ and $K_0$ are the modified Bessel functions, $c_k$ are arbitrary constants. The
corresponding eigenvalue $\varepsilon_0$ is equal to $c_j$, i.e.,
\[\varepsilon_0=-\frac{1}{4(j+1)^2}.\]

 The solution $\Phi^n_{j,k}$ corresponding to $n^{th}$ exited state
and the corresponding eigenvalue $\varepsilon_n$
 are:
 \begin{gather}\label{psin}\Phi^n_{j,k}=
a_{j}^+a_{j+1}^+ \cdots
a_{j+n-1}^+\Phi^0_{j+n,k},\quad\varepsilon_n=-\frac{1}{4(j+n+1)^2}.
\end{gather} The related energy value in (\ref{EP}) is given by the
following equation:
\begin{gather}\la{el} E=-\frac{m\alpha^2}{2N^2}, \end{gather}where
\begin{gather}\label{ela}N=n+j+1,\quad n=0, 1, 2,\dots\end{gather}

The energy levels (\ref{el})  depend on the main quantum number $N$ (\ref{ela}) which can take the same value for different pairs of $n$ and $j$.  Namely, for a fixed $N$ there are $N-\frac12$ of such pairs. In addition, the quantum number $k$ which labels the eigenvectors in (\ref{rep}) is not present in (\ref{el}) and (\ref{ela}). Thus like in the HA problem the energy  levels are
highly degenerated, and this degeneration is caused by the hidden
symmetry w.r.t. group O(4).

  To end this section we remind that the
radial component of the HA Hamiltonian (\ref{H1}) is shape invariant
also and can be factorized like (\ref{fac}) where $j\to l$, $W\to
\frac{q}{2(l+1)}+\frac{l+1}r$ and $l$ is the quantum number labeling
eigenvalues of angular momentum $L$. In other words, our model
succeeds both hidden symmetry and supersymmetry of the HA.
\section{Relativistic system}
Let us show that eigenvalue problem (\ref{ep}) admits a relativistic
formulation. To demonstrate that we start with the Dirac equation
for a neutral particle  having non-trivial dipole moments:
\begin{gather}\left(\gamma^\mu p_\mu-m-\alpha
\sigma^{\mu\nu}F_{\mu\nu}+{\tilde \alpha}\gamma_5\gamma^\mu
F_\mu\right)\Psi=0.\la{di}\end{gather}

In addition to the standard Pauli term $\alpha \sigma^{\mu\nu}F_{\mu\nu}$
Eq. (\ref{di}) includes additional term ${\tilde
\alpha}\gamma_5\gamma^\mu F_\mu$ with an external pseudovector field
$F_\mu$. For convenience the following realization of Dirac matrices
will be used:
\begin{gather}\la{dm}\gamma_0=\begin{pmatrix}0&I\\I&0\end{pmatrix},\quad
\gamma_5=\begin{pmatrix}0&-I\\I&0\end{pmatrix},\quad
\gamma_a=\begin{pmatrix}\ri\sigma_a&0\\0&-\ri \sigma_a\end{pmatrix}
\end{gather}
where $I$ and $0$ are the $2\times2$ unit and zero matrices
correspondingly, $a=1, 2, 3$.

Choosing in (\ref{di})
\begin{gather}\la{F}F_{ab}=F_0=0, \
F_a=\frac{\alpha}{\tilde \alpha}F_{0a}=\frac{x_a}{x^2}\end{gather} and
representing $\Psi$ as:
\begin{gather}\la{ppsi}\Psi=\exp(-iEx_0)\begin{pmatrix}\psi({\bf x})\\
\xi({\bf x})\end{pmatrix}\end{gather}  with two-component functions $\psi$ and
$\xi$, we come to the following system:
\begin{gather}\la{sys1}(\ri{\mbox{\boldmath $\sigma$}}\cdot{\bf
p}-m)\psi+E\xi=0,\\ \la{sys2}\left(E-2\alpha\frac{{\mbox{\boldmath
$\sigma$}}\cdot {\bf x}}{x^2}\right)\psi-(m+\ri{\mbox{\boldmath
$\sigma$}}\cdot{\bf p})\xi=0.\end{gather}

Solving (\ref{sys1}) for $\xi$ and substituting this solution into
(\ref{sys2}) we come to the eigenvalue problem (\ref{ep}) where
${\bf r}=2\alpha E\bf x$ and $\varepsilon=\frac{E^2-m^2}{4\alpha^2E^2}$. Thus,
in accordance with (\ref{psin}) the relativistic energies are
discrete also and can be given by the following formula:
\begin{gather}\la{EL}E^2=\frac{m^2}{1+\frac{\alpha^2}{N^2}}\end{gather}
with $N$ being a natural number which can be represented in the form
 (\ref{ela}). The corresponding (non-normalized) wave
functions are given by Eq. (\ref{ppsi}) where $\psi({\bf x})$
is defined in (\ref{SS}), (\ref{phi}), (\ref{gs}), (\ref{psin}) and
$\xi({\bf x})$ can be found from (\ref{sys1}).

 Notice that for small coupling constants $\alpha$ the positive energy values (\ref{EL}) are reduced to the nonrelativistic form (\ref{el}) up to the constant internal energy term $m$ and terms of order $\alpha^4$.

 \section{Superintegrable and supersymmetric system with spin-orbit interaction}

 In very recent preprint \cite{wint3} one more QM model with a hidden symmetry w.r.t. group O(4) is presented. The Hamiltonian of this model looks as follows:
\begin{gather}H=\frac{p^2}{2m}+\hat V-\frac1{8x^2}\la{W1}\end{gather}
where
\begin{gather}\la{W2}\hat V=\frac1{2mx^2}\left({\mbox{\boldmath $\sigma$} \cdot \bf
L}+1\right)-\frac{\alpha}{x}.\end{gather}

In contrast with (\ref{H2}) Hamiltonian (\ref{W1}) does not include the dipole interaction term but involves a scalar potential together with the spin-orbit interaction. Nevertheless, it also commutes with the total orbital momentum and generalized Runge-Lenz vector (\ref{J}) where $\hat V$ is given by equation (\ref{W2}) (in paper \cite{wint2} another representation for $\hat R$ is used which is equivalent to (\ref{J}), (\ref{W2})). This result was predictable since Hamiltonian (\ref{W1}) can be reduced to the direct sum of two Hamiltonians of HA via the gauge transformation \cite{wint}.

It can be proven by direct verification, that
 operators (\ref{J}), (\ref{W2}) and (\ref{W1}) satisfy commutation relations (\ref{CR})  and so they generate a hidden symmetry with respect to group O(4). Thus the QM system with Hamiltonian (\ref{W1}) admits a generalized Fock symmetry.

 Hamiltonian (\ref{W1}) succeeds also one more symmetry of the systems considered in the above, i.e., the shape invariance. Indeed, starting with (\ref{W1}) and repeating  all steps following equation (\ref{ep}), one comes to radial equation (\ref{rep}) where
 \begin{gather}V_j=\frac{j(j+1)}{r^2}-\frac{1}{r}.\label{W3}\end{gather}
Here and in the next equation each term in the r.h.s. is multiplied
by a $2\times2$ identity matrix.

 Like (\ref{pot}), potential (\ref{W3})
  is shape invariant. The related superpotential is given by the following formula:
  \begin{gather} W=\frac{1}{2(j+1)}+\frac{j+1}r\label{W4}\end{gather}
  while eigenvalues of Hamiltonian (\ref{W1}) are given by equation (\ref{el}) where $n$ and $j$ are positive integers and  half integers correspondingly. The corresponding wave function is easily calculated in complete analogy with section 4. It is given by equation (\ref{SS}) where $\psi_{j,\lambda,k}\to\psi^n_{j,\lambda,k}$ and
  \begin{gather}\psi^n_{j,\lambda,k}=c_{\lambda k}y^{j+1}\exp\left(-\frac{y}2\right)L_n^{2j+1}(y).\end{gather}
  Here $y=\frac{r\alpha^2}{n+j+1}$, $L_n^{2j+1}(y)$ are Laguerre polynomials
  and $c_{\lambda k}$ are constants satisfying the normalizing condition $\sum_{\lambda, k} c_{\lambda k}c^*_{\lambda k}=1$.

Effective potential (\ref{W3}) and superpotential (\ref{W4}) are similar to ones for the HA. However, there are some differences, namely:
\begin{itemize}
\item Potential (\ref{W3}) includes quantum number $j$ which takes half-integer values while the potential of the HA depends on orbital quantum number $l$ which is integer;
\item In contrast with the HA, potential (\ref{W3}) is a matrix. More exactly, it is a direct sum of two scalar potentials. That leads to the additional two-fold degeneration of the energy spectrum of Hamiltonian (\ref{W1}). This degeneration is caused by the additional integral of motion
\begin{gather}C=\frac1{\alpha}{\bf J}\cdot{\hat {\bf R}}= \frac{\mbox{\boldmath $\sigma$} \cdot \bf
x}{2x}\la{W5}\end{gather} whose eigenvalues are not included into the spectrum formula (\ref{el}).
\item Matrix (\ref{W5}) is proportional to the Casimir operator $\hat C={\bf J}\cdot\hat {\bf R}$ of group O(4), and its eigenvalues are equal to $\pm \frac12$. For the HA this Casimir operator is trivial.
\item The additional symmetry operator (\ref{W5}) extends the number of algebraically independent constants of motion (including the Hamiltonian) to 6. This number is maximal for a 3d system with
    the additional (spin) degree of freedom, thus the discussed system is maximally superintegrable.

\end{itemize}

The matrix (\ref{W5}) and many other integrals of motion for Hamiltonian (\ref{W1}) were represented in paper \cite{wint}. All of them are algebraic functions of the basic symmetry operators (\ref{J}) where $\hat V$ is the matrix given in (\ref{W2}).

\section{Discussion}

The model Hamiltonian (\ref{H2}) represents a new exactly solvable
QM system which admits extended symmetries. Like the HA this system
admits the hidden (Fock) symmetry w.r.t. group O(4) whose generators
are the total angular momentum and Runge-Lenz vector. In addition
(and again like the HA) this system is supersymmetric and can be
easily solved using tools of SUSY QM, see Section 4.
Moreover, this is a fermionic system with spin $\frac12$ while the
non-relativistic model of HA ignores the spin of electron.

A new feature of the supersymmetry of Hamiltonian (\ref{H2}) in
comparison with the HA is that it is realized in terms of matrix
superpotentials. Such (one dimensional) superpotentials had been
classified in papers \cite{NK1} and \cite{NK2}, and it was very
inspiring for us to search for multidimensional matrix systems which
can be solved using their matrix supersymmetry in separated
variables. As a result Hamiltonian (\ref{H2}) has been
discovered. The other results of such search are presented in papers
\cite{N1} and \cite{N2}.

Operator (\ref{H2}) can be interpreted as the Hamiltonian of a neutral
particle with spin 1/2 which has a non-trivial  dipole moment.
Matrix potential $\alpha\frac{{\mbox{\boldmath $\sigma$}}\cdot{\bf
x}}{x^2}=\alpha{{\mbox{\boldmath $\sigma$}}\cdot{\bf
E}}$ represents a Pauli type interaction with the external
vector field
\begin{gather}\la{la}{\bf E}\sim\frac{\bf x}{x^2}.\end{gather}
The latter can be interpreted as the electric field. Moreover,
such field can be realized experimentally at least on finite
interval $a<x<b, a>0$, see, e.g., problem 1018 in \cite{book}. Notice
that functions (\ref{la}) also solve nonlinear Maxwell equations
including additional vector field \cite{NKu}. They also solve equations of axion electrodynamics
\cite{NKuku}. The same is true for
fields (\ref{F}) involved into relativistic equation (\ref{di}).

Thus the QM systems whose Hamiltonians are given by equations  (\ref{H2}) and (\ref{W1})
posses all symmetries and supersymmetries  admitted by the  HA. However
these symmetries are generalized by introducing the spin. In
addition, there exist a relativistic counterpart of system (\ref{EP})
given by Eqs. (\ref{di}) and (\ref{F}) which is exactly
solvable too.

The presented non relativistic system (\ref{EP}) can be treated as a 3d generalization of the solvable planar system proposed by Pronko and Stroganov \cite{Pron}. It would be interesting to study its possible generalizations in multidimensional spaces. One more interesting task  is to construct solvable systems with Fock symmetry for particles with arbitrary spin. This work is in progress.

It is well known that there exists a tight coupling between the hidden symmetry of the HA and its supersymmetry \cite{Mo}, \cite{mota}. The same is true for the Pronko-Stroganov problem \cite{mota2}. A contemporary discussion  of relations between supersymmetry and superintegrability can be found in \cite{Ra}.

All systems discussed in the present paper are both supersymmetric and maximally superintegrable. Surely, these combinations of symmetries are not accidental, and it would be interesting to extend the results \cite{mota}, \cite{mota2} to the 3d matrix systems presented here.
\begin{center}
Acknowledgments
\end{center}
I am indebted to Professor P. Winternitz  for an inspiring discussion about superintegrability in systems with spin.

\end{document}